\documentclass[12pt,twoside]{article}
\usepackage{fleqn,espcrc1}

\def\beq{\begin{equation}}
\def\eeq{\end{equation}}
\def\bea{\begin{eqnarray}}
\def\eea{\end{eqnarray}}
\def\bq{\begin{quote}}
\def\eq{\end{quote}}

\def\nnb{\nonumber}
\def\ga{\left(}
\def\dr{\right)}

\def\rar{\rightarrow}
\def\lrar{\Longrightarrow}

\def\nnb{\nonumber}
\def\la{\langle}
\def\ra{\rangle}
\def\nin{\noindent}
\def\ba{\begin{array}}
\def\ea{\end{array}}

\def\bl{\bullet}

\def\als{\alpha_s}

\def\g2{ \la\alpha_s G^2 \ra}
\def\g3{g^3f_{abc}\la G^aG^bG^c \ra}
\def\g4{\la\als^2G^4\ra}

\newcommand{\AmS}{{\protect\the\textfont2
  A\kern-.1667em\lower.5ex\hbox{M}\kern-.125emS}}

\title{Gluonia, Scalar and Hybrid Mesons in QCD\thanks{Plenary talk
given at Hadron 99 (Beijing 24-28th August 1999)}}

\author{Stephan Narison\address{
Laboratoire de Physique Math\'ematique,
Universit\'e de Montpellier II
Place Eug\`ene Bataillon,
34095 - Montpellier Cedex 05, France.
E-mail:
narison@lpm.univ-montp2.fr}}%
        
\begin{document}

% typeset front matter
\maketitle

\begin{abstract}
For some experimental guidelines of the next millenium, I review the determinations of the
masses, decays and mixings of the gluonia, scalar and hybrid
mesons from QCD spectral sum rules and low-energy theorems, and compare them with the
lattice.
\end{abstract}

\section{INTRODUCTION}
Since the discovery of QCD, it has been emphasized  \cite{GM} that exotic mesons beyond the
standard octet, exist as a consequence of the non-perturbative aspects of
quantum chromodynamics (QCD). Since the understanding of the nature of the $\eta'$
\cite{U1}, a large amount of theoretical efforts have been furnished in the past
and pursued at present for predicting the spectra of the exotics using
different QCD-like models such as the flux tube \cite{REVF} , the bags
\cite{REVB}, the quark \cite{REVP} and constituent gluon \cite{REVG} models . In this talk,
I shall review the present status of the predictions from the QCD spectral sum rules (QSSR)
\`a la SVZ
\cite{SVZ} (for a review, see e.g.: \cite{SNB}) and from some low-energy theorems based on
Ward identities, which I
will compare with the lattice results. 

\section{QCD SPECTRAL SUM RULES (QSSR)}
\subsection{Description of the method}

Since its discovery in 79, QSSR has proved to be a
powerful method for understanding the hadronic properties in terms of the
fundamental QCD parameters such as the QCD coupling $\alpha_s$, the (running)
quark masses and the quark and/or gluon QCD vacuum condensates.
The description of the method has been often discussed in the literature,
where a pedagogical introduction can be, for instance, found in the book \cite{SNB}. In
practice (like also the lattice), one starts the analysis from the two-point correlator:
\beq
\psi_H(q^2) \equiv i \int d^4x ~e^{iqx} \
\la 0\vert {\cal T}
J_H(x)
\ga J_H(0)\dr ^\dagger \vert 0 \ra ~,
\eeq
built from the hadronic local currents $J_H(x)$, which select some specific quantum numbers.
However, unlike the lattice which evaluates the correlator in the Minkowski space-time,
one exploits, in the sum rule approaches, the analyticity property of the
correlator which obeys the well-known K\"allen--Lehmann dispersion relation:
\beq
\psi_H (q^2) = 
\int_{0}^{\infty} \frac{dt}{t-q^2-i\epsilon}
~\frac{1}{\pi}~\mbox{Im}  \psi_H(t) ~ + ...,
\eeq
where ... represent subtraction points, which are
polynomials in the $q^2$-variable. In this way, the $sum~rule$
expresses in a clear way the {\it duality} between the integral involving the 
spectral function Im$ \psi_H(t)$ (which can be measured experimentally), 
and the full correlator $\psi_H(q^2)$. The latter 
can be calculated directly in the
QCD Euclidean space-time using perturbation theory (provided  that
$-q^2+m^2$ ($m$ being the quark mass) is much greater than $\Lambda^2$), and the Wilson
expansion in terms of the increasing dimensions of the quark and/or gluon condensates which
 simulate the non-perturbative effects of QCD. 

\subsection{Beyond the usual SVZ expansion}

Using the Operator Product Expansion (OPE) \cite{SVZ}, the two-point
correlator reads:
\beq
\psi_H(q^2)
\simeq \sum_{D=0,2,4,...}\frac{1}{\ga -q^2 \dr^{D/2}} 
\sum_{dim O=D} C(q^2,\nu)\la {\cal O}(\nu)\ra~,
\eeq
where $\nu$ is an arbitrary scale that separates the long- and
short-distance dynamics; $C$ are the Wilson coefficients calculable
in perturbative QCD by means of Feynman diagrams techniques; $\la {\cal O}(\nu)\ra$
are the quark and/or gluon condensates of dimension $D$.
In this paper, we work in the massless quark limit. Then, one may expect
the absence of the terms of dimension 2 due to gauge invariance. However, it has been
emphasized recently \cite{ZAK} that  the resummation of the large order terms of the
perturbative series, and the effects of the higher dimension condensates due e.g. to instantons, can
be mimiced by the effect of a tachyonic gluon mass, which might be understood
from the short distance linear part of the QCD potential. The strength of
this short distance mass has been estimated from the $e^+e^-$ data to be
\cite{SNI,CNZ}:
$
\frac{\alpha_s}{\pi}\lambda^2\simeq -(0.06\sim 0.07) ~\rm{ GeV}^2,
$
which leads to the value of the square of the (short distance) string tension:
$
\sigma \simeq -\frac{2}{3}{\alpha_s}\lambda^2\simeq [(400\pm 20)~\rm{ MeV}]^2
$
in an (unexpected) good agreement with the lattice result \cite{TEPER} of about
$[(440\pm 38)~\rm{ MeV}]^2$.
The strengths of the vacuum condensates having dimensions $D\leq 6$ are also under
good control: namely $2m\la\bar qq\ra =-m^2_\pi f^2_\pi$ from pion PCAC,
$\la\alpha_s G^2\ra =(0.07\pm 0.01)$ GeV$^2$ from $e^+e^-\rar I=1$ data \cite{SNI} and from 
the heavy quark mass-splittings \cite{SNH}, $\alpha_s  \la\bar qq\ra^2\simeq
5.8 \times 10^{-4}$ GeV$^6$ \cite{SNI}, and $g^3\la G^3\ra\approx$ 1.2 GeV$^2\la\alpha_s G^2\ra$ from
dilute gaz instantons \cite{NSVZ}.
\subsection{Spectral function}
%In the absence of the complete data, 
The spectral function is often parametrized
using the ``na\"{\i}ve" duality ansatz:
\beq
\frac{1}{\pi}~\mbox{Im}  \psi_H(t)\simeq 2M_H^{2n}f_H^2 \delta (t-M_H^2)+ \rm{``QCD
~continuum"}
\times \theta(t-t_c)~, 
\eeq
which has been tested \cite{SNB} using $e^+e^-,~\tau$-decay data, to give a good description of the
spectral integral in the sum rule analysis; $f_H$ (analogue to $f_\pi$) is the
the hadron's coupling to the current ; $2n$ is the dimension of the
correlator, while $t_c$ is the QCD continuum's threshold. 

\subsection{Form of the sum rules and optimization procedure}
Among the different sum rules discussed in the literature \cite{SNB} (Finite
Energy Sum rule (FESR) \cite{RAFAEL}, $\tau$-like sum rules \cite{BNP},...), we shall mainly be
concerned here with:\\
$\bullet$ The exponential Laplace unsubtracted sum rule (USR)
and its ratio:
\beq\label{usr}
{\cal L}_n(\tau)
= \int_{0}^{\infty} {dt}~t^n~\mbox{exp}(-t\tau)
~\frac{1}{\pi}~\mbox{Im} \psi_H(t)~,~~~~~~~~~~{\cal R}_{n} \equiv -\frac{d}{d\tau} \log {{\cal
L}_n}~,~~~~~~~(n\geq 0)~;
\eeq
$\bullet$ The subtracted sum rule (SSR):
\beq\label{ssr}
{\cal L}_{-1}(\tau)
= \int_{0}^{\infty} \frac{dt}{t}~\mbox{exp}(-t\tau)
~\frac{1}{\pi}~\mbox{Im} \psi_H(t) +\psi_H(0)~.
\eeq
The advantage of the Laplace sum 
rules with respect to the previous dispersion relation is the
presence of the exponential weight factor, which enhances the 
contribution of the lowest resonance and low-energy region
accessible experimentally. For the QCD side, this procedure has
eliminated the ambiguity carried by subtraction constants,
arbitrary polynomial
in $q^2$, and has improved the convergence of
the OPE by the presence of the factorial dumping factor for each
condensates of given dimensions. 
The ratio of the sum rules is a useful quantity to work with,
 in the determination of the resonance mass, as it is equal to the 
meson mass squared, in the usual duality ansatz parametrization.
As one can notice, there are ``a priori" two free external parameters $(\tau,
t_c)$ in the analysis. The optimized result will be (in principle) insensitive
to their variations. In some cases, the $t_c$-stability is not reached due to the
too na\"{\i}ve parametrization of the spectral function. One can either fixed the 
$t_c$-values by the help of FESR (local duality) or improve the
parametrization of the spectral function by introducing threshold effects fixed by
chiral perturbation theory, ..., in order to restore the $t_c$-stability of the
results. The results discussed below satisfy these stability criteria.
\section{UNMIXED GLUONIA CURRENTS, MASSES AND COUPLINGS}
\subsection{The currents}
In this paper, we shall consider the lowest-dimension gluonic currents
that can be built from the gluon fields $G^a_{\alpha\beta}$ and which are gauge-invariant:
\bea
\theta_{\mu}^{\mu}&=& \beta(\alpha_s) G_{\alpha\beta}G^{\alpha\beta}+\sum_{u,d,s}m_q\bar
qq~,~~~~~
\theta_{\mu\nu}=-G_{\mu}^{\alpha}G_{\nu\alpha}+\frac{1}{4}g_{\mu\nu}
G_{\alpha\beta}G^{\alpha\beta}, \nnb\\
\partial_\mu A^\mu(x)&=&\ga \frac{\alpha_s}{8\pi}\dr
\mbox{tr}~G_{\alpha\beta}
\tilde{G}^{\alpha\beta}+\sum_{u,d,s}m_q\bar
q(i\gamma_5)q~,~~~~~~
J_3=g f_{abc} G^a_{\alpha\beta}G^b_{\beta,\gamma}G^c_{\gamma\alpha}~,
\eea
where the sum over colour is understood; $\theta_{\mu}^{\mu}$ is the
trace of the energy-momentum tensor $\theta_{\mu\nu}$; $\partial_\mu A^\mu(x)$ is the
$U(1)_A$ anomaly; $m_q$ is the light quark mass and $\beta$ is the $\beta$-function.  
\subsection{Masses and couplings}
The unmixed gluonia masses from the unsubtracted QCD Spectral Sum Rules
(USR) \cite{SNG} are compared in Table 1 with the ones from the lattice
\cite{TEPER,LATT} in the quenched approximation, where we use the
conservative guessed estimate of about
15\% for the different lattice systematic errors  (separation  of the lowest ground
states from the radial excitations, which are expected to be nearby as
indicated by the sum rule analysis; discretisation; quenched
approximation,...). 
\begin{table}[hbt]
\caption{ Unmixed gluonia masses and couplings from QSSR \cite{SNG} compared with the lattice.  } 
\begin{center}
\begin{tabular}[h]{ccccccc}
\hline 
%\hline
% & & & & & &\\
$J^{PC}$& Name&\multicolumn{3}{l} {Mass [GeV]
}& $\sqrt{t_c}$ [GeV]&$f_H$ [MeV]\\
% & & & & & &\\
\cline{3-5}
% & & & & & &\\
&&Estimate& Upper Bound&Lattice \cite{TEPER,LATT} &&  \\
%&&&&& \\
%&&&&&&\\
\hline 
%\hline
%& & & & & & \\
 $0^{++}$&$G$&$ 1.5\pm 0.2$& $2.16\pm 0.22$&$1.60\pm
0.16$&$2.1$&$390\pm 145$\\
%& & & & & & \\
&$3G$& 3.1&3.7&&3.4&62\\
%&&&&&&\\
$2^{++}$&$T$&$2.0\pm 0.1$&$2.7\pm 0.4$&$2.26\pm
0.22$&$2.2$&$80\pm 14$ \\ 
%&&&&&&\\
$0^{-+}$&$P$&$2.05\pm 0.19$&$2.34\pm 0.42$&$2.19\pm
0.32$&$2.2$&$8
\sim17$ \\
%& &&&&&\\
\hline 
%\hline
\end{tabular}
\end{center}
\end{table}
One can notice an excellent agreement between the USR 
 and the lattice quenched results, which the mass hierarchy:
$
 M_{0^{++}}\leq M_{0^{-+}}\approx M_{2^{++}}$,  expected from some QCD inequalities \cite{WEST}.
However, this is not the whole story !
\section{PSEUDOSCALAR GLUONIA}
\subsection{Testing the nature of the $E/\iota$}
We test the gluonic nature of the $E/\iota$ by determining its decay constant
$f_\iota$, from a saturation of the USR (Eq.\ref{usr}) and SSR (Eq.\ref{ssr}) pseusdoscalar sum rules by the
$\eta'$,
$E/\iota$ and the gluonium $P$. One obtains \cite{SNG} a value of $f_\iota$ consistent with
zero, in agreement with the estimate from the $J/\psi\rar \gamma+E/\iota$ decays
($f_\iota\approx 7$ MeV),
and smaller than $f_\eta'\approx 30$ MeV \cite{SHORE} and $f_P\approx (8\sim 17)$ MeV. 
The quarkonium-gluonium {\it mass mixing angle} is determined to be small ($\theta_P\approx 12^0$), from the
off-diagonal two-point correlator \cite{PAK,SNB} (see also \cite{LIPKIN}):
\beq\label{off}
\psi_{qg}(q^2) \equiv i \int d^4x ~e^{iqx}
\la 0\vert {\cal T}
\ga\frac{\alpha_s}{8\pi}\dr
\mbox{tr}~G_{\alpha\beta}
\tilde{G}^{\alpha\beta}(x)\sum_{u,d,s}m_q\bar
q(i\gamma_5)q~(0) \vert 0 \ra~ ,
\eeq
 which, then, suggests a small mass shift of the physical states after mixing.\\ {\it We
may conclude that the $E/\iota$ is likely the radial excitation of the $\eta$ or/and $\eta'$}.
 \subsection{Radiative decays of the pseudoscalar gluonium P}
Using $|\theta_P|\approx 12^0$, and the OZI rule, one can predict \cite{PAK,SNB,SNG}: 
\beq
\Gamma(P\rar\gamma\gamma)\approx (1.3\pm 0.1)~\rm{keV}~,~~~~~~~~~~
\Gamma(P\rar\rho\gamma)\approx (.3\pm 0.1)~\rm{keV}~,
\eeq
where the errors are probably underestimated. This result is testable at BES.
\section{TENSOR GLUONIA}
\subsection{Spectrum}
From Table 1, the lowest ground state and the radial
excitations of about $\sqrt{t_c}$ are almost degenerated in masses, which suggests a
rich population of the
$2^{++}$ gluonia around 2 GeV. Though the $\zeta(2.2)$ is a good gluonium candidate \cite{BES}, it may not be
the lowest ground state. A complete analysis needs systematic scannings of the region above 1.9 GeV and
further tests of the old BNL candidates $g_T$.
\subsection{Quarkonium-gluonium mass mixing}
A QSSR analysis of the tensor correlator \cite{BRAMON,SNB}, similar to the one in
Eq.(\ref{off}), leads to a small quarkonium-gluonium {\it mass mixing} angle of about $-10^0$.
\subsection{Tensor $T$ decays}
One starts from the universality of the vertex form factor \cite{SHIF}:
\beq
\la\pi(p)|\theta_{\mu\nu}|\pi(p')\ra\simeq {\rm (Lorentz~structure)}\times 1,
~~~{\rm at}~~~ q^2\equiv (p-p')^2=0,
\eeq
and write a dispersion relation. Using the $f_2\rar\pi\pi$ data, one can deduce \cite{SNB,BRAMON}:
\beq
\Gamma(T\rar\pi\pi+KK+\eta\eta)\leq (119\pm 36)~{\rm MeV})~,~~~~~
\Gamma(T\rar\pi\pi)\approx 10~{\rm MeV}\leq 70~{\rm MeV}~,~\eeq
in agreement with present data \cite{BUGG}. A non-relativistic relation between the $0^{++}$ and the $2^{++}$
wave functions gives the width: \beq\Gamma(T\rar\gamma\gamma)\approx 0.06~{\rm keV}~.\eeq
%\eeq
\section{UNMIXED SCALAR GLUONIA}
\subsection{The need for a low mass $\sigma_B$ from the sum rules}
Using the mass and coupling of the scalar gluonium $G$ in Table 1 from
the USR (Eq.\ref{usr}), into the SSR (Eq.\ref{ssr}) sum rules, where \cite{NSVZ} $\psi_s(0)\simeq
-16(\beta_1/\pi)\la
\alpha_s G^2\ra$, one can notice \cite{VEN,SNG} that one needs a low mass resonance $\sigma_B$ for a consistency of
the two sum rules. Using $M_{\sigma_B}\simeq 1$ GeV, one gets
\cite{VEN,SNG}:
$
f_{\sigma_B}\approx 1 ~\rm{GeV}
$, which is larger than $f_G\simeq$ .4 GeV.
\subsection{Low-energy theorems (LET) for the couplings to meson pairs} 
In order to estimate the couplings of the gluonium to meson pairs, we use some sets of low-energy theorems (LET)
based on Ward identities for the vertex:
\beq
V(q^2\equiv (p-p')^2=0)\equiv \la H(p)|\theta_{\mu}^{\mu}|H(p')\ra\simeq 2m^2_H~,~~ ~~~~\rm{and}~~~~~~ V'(0)=1~,
\eeq
and write the vertex in a dispersive form. $H$ can be a Goldstone boson ($\pi,K,\eta_8$), a $\eta_1$-
$U(1)_A$-singlet~, or a $\sigma_B$. Then, one obtains the sum rules for the hadronic couplings:
\beq
\frac{1}{4}\sum_{\sigma_B, \sigma'_B, G}g_{SHH}\sqrt{2}f_S\approx 2M^2_H~,~~~~~~~~~~~~~~~~~~~~~~~
\frac{1}{4}\sum_{\sigma_B, \sigma'_B, G}g_{SHH}\sqrt{2}f_S/M^2_S\simeq 1~.
\eeq
%\begin{itemize}
$\bl$ Neglecting, to a first approximation the $G$-contribution, the $\sigma_B$ and $\sigma'_B$ widths to
$\pi\pi,~KK,...$
 (we take $M_{\sigma'}\approx
1.37$ GeV as an illustration) are \cite{SNG}:
\beq
\Gamma(\sigma_B\rar\pi\pi)\approx 0.8~{\rm GeV}~,~~~~~~~~~~~~~~~~~~~~~~
\Gamma(\sigma'_B\rar\pi\pi)\approx 2~{\rm GeV}~,
\eeq
which suggests a huge OZI violation and seriously questions the validity of the
lattice results in the quenched approximation. Similar conclusions have been reached in
\cite{MONT,MINK,ANIS}. For testing the above
result, one should evaluate on the lattice, the decay mixing 3-point function $V(0)$ responsible for such decays
using dynamical fermions. \\
$\bl$ The previous LET 
implies the {\it characteristic gluonium decay} (we use $M_G\approx 1.5$ GeV and assume a
G-dominance) \cite{VEN,SNB}:
\beq
\Gamma(G\rar\eta\eta')\approx (5-10)~{\rm MeV}~,~~~~~~
\frac{\Gamma(G\rar\eta\eta)}{\Gamma(G\rar\eta\eta')}\approx 0.22 ~~~:
~~~g_{G\eta\eta}\simeq \sin\theta_P~g_{G\eta\eta'}~.
\eeq
$\bl$ { Assuming that the $G$ decay into $4\pi^0$ occurs through $\sigma_B\sigma_B$},
and using the data for $f_0(1.37) \rar (4\pi^0)_S$, one obtains \cite{VEN,SNB}:
\beq\Gamma(G\rar\sigma_B\sigma_B\rar 4\pi)\approx (60-140)~{\rm MeV}~.
\eeq
%\end{itemize}
\subsection{$\gamma\gamma$ widths and $J/\psi\rar\gamma S$  radiative decays}
These widths can be estimated from the quark box or anomaly diagrams \cite{VEN,SNB}. 
The $ \gamma\gamma$ widths
of the $\sigma, \sigma'$ and $G$ are much smaller (factor 2 to 5) than
$\Gamma(\eta'\rar\gamma\gamma) \simeq$ 4 keV, while
$B (J/\psi\rar \gamma$ $\sigma, \sigma'$ and $G$) is about 
10 times smaller than 
$B(J/\psi\rar \gamma \eta')\approx 4~
10^{-3}$. These are typical values of gluonia widths and production rates \cite{CHAN}.
\section{UNMIXED SCALAR QUARKONIA}
\subsection{The $a_0(980)$}
The  $a_0(980)$ is the most natural meson candidate associated to the divergence of the vector
current:
$
\partial_\mu V^\mu (x)\equiv (m_u-m_d) \bar u(i\gamma_5) d.
$
Previous different sum rule analysis of the associated two-point
correlator gives \cite{SNB}:
$
M_{a_0}\simeq 1~\rm{ GeV}$ and the conservative range $ f_{a_0}\simeq (0.5-1.6) ~{\mbox
MeV}~~(f_\pi=93~{\mbox MeV}), $ 
in agreement with the value 1.8 MeV from a hadronic kaon tadpole mass difference approach plus
a $a_0$ dominance of the $K\bar K$ form factor, and includes the recent sum rule determination
\cite{MALT}. A
3-point function sum rule analysis gives the widths \cite{BRAMON2,SN4,SNB}:
\beq
\Gamma (a_0\rar\eta\pi) \simeq 37~ {\rm MeV}~,~~~~~~~~~~~~~~~~~~~~~~~~\Gamma (a_0\rar\gamma\gamma)
\simeq (0.3-1.5)~{\rm keV}~,
\eeq
while from $SU(3)$ symmetry, we expect to have:
$
g_{a_0 K^+K^0}\simeq \sqrt{
\frac{3}{2}} g_{a_0\eta\pi}.
$
Analogous sum rule  analysis in the four-quark scheme \cite{SN4,SNB} gives similar values of the masses
and hadronic couplings but implies a too small value of the $\gamma\gamma$ width
due to the standard QCD loop-diagram factor suppressions.
The $(\bar uu-\bar dd)$ quark assignement for the $a_0(980)$ is supported by present data
and alternative approaches \cite{MONT}. 
\subsection{The isoscalar partner $S_2\equiv \bar uu+\bar dd$ of the $a_0(980)$}
Analogous analysis of the corresponding 2-point correlator gives $M_{S_2}\approx M_{a_0}$
as expected from a good $SU(2)$ symmetry, while its widths are estimated to be \cite{SNG,BRAMON2}:
\beq
\Gamma (S_2\rar\pi^+\pi^-) \simeq 120~ {\rm MeV}~,~~~~~~~~~~~~~~~
\Gamma (S_2\rar\gamma\gamma)\simeq \frac{25}{9}\Gamma (a_0\rar\gamma\gamma) \simeq
0.7~{\rm keV}~.
\eeq
\subsection{The $K^*_0(1430)\equiv \bar us$ and $S_3\equiv \bar ss$ states}
The $K^*_0(1430)$ is the natural partner of the $a_0(980)$, where their mass shift is due to $SU(3)$ breakings
 \cite{SNB}.  An analysis of the $S_3$ over the $K^*_0$ 2-point
functions gives
\cite{SNG}:
\beq
M_{S_3}/M_{K^*_0}\simeq 1.03\pm 0.02 ~~\lrar~~~M_{S_3}\simeq 1474~{\rm MeV}~,~~~~~~~~f_{S_3}\simeq (43\pm
19)~\rm{MeV}~,\eeq in agreement with the lattice result \cite{LEE}, while the 3-point function leads to
\cite{SNG}:
\beq
\Gamma (S_3\rar K^+K^-)\simeq (73\pm 27)~{\rm MeV}~,~~~~~~~~~~~~
 \Gamma (S_3\rar\gamma\gamma) 
\simeq 0.4~{\rm keV}~.
\eeq
In the usual sum rule approach, and in the absence of large violations of the OPE at the sum rule stability
points, one expects a small mixing between the $S_2$ and $S_3$ mesons before the mixing with the gluonium
$\sigma_B$. 
\subsection{Radial excitations}
The propreties of the radial excitations cannot be obtained accurately from the sum rule
approach, as they are part of the QCD continuum which effects are minimized in the analysis.
However, as a crude approximation and using the sum rule results from the well-known channels ($\rho$,...),
one may expect that the value of $\sqrt{t_c}$ can localize approximately the position of the first radial
excitations. Using this result and some standard phenomenological arguments on the estimate of the
couplings, one may expect \cite{SNG}:
\bea
M_{S'_2}\approx 1.3~\rm{GeV},~~\Gamma (S'_2\rar \pi^+\pi^-)\approx (300\pm
150)~{\rm MeV},~~
 \Gamma (S'_2\rar\gamma\gamma)\approx (4\pm 2)~{\rm keV}~,\nnb\\
M_{S'_3}\approx  1.7~{\rm GeV},~~
\Gamma (S'_3\rar K^+K^-)\approx (112\pm 50)~{\rm
MeV},~~
 \Gamma (S'_3\rar\gamma\gamma)\approx (1\pm .5)~{\rm keV}.
\eea
\subsection{We conclude that:}
%The analysis of the unmixed scalar sector shows that:\\
%\begin{itemize}
\nin
$\bullet$ Unmixed scalar quarkonia ground states are not wide, which excludes the interpretation
of the low mass broad $\sigma$ for being an ordinary $\bar qq$ state.\\
$\bullet$ There can be many states in the region  around 1.3 GeV ($\sigma', ~S_3, ~S'_2$),which 
should mix non-trivially in order to give the observed $f_0(1.37)$ and $f_0(1.5)$
states (see next sections).\\
$\bullet$ The $f_J(1.7)$ seen to decay mainly into $\bar KK$ \cite{SING}, if it is
confirmed to be a $0^{++}$ state, can be {\it the first radial excitation of the $S_3\equiv\bar ss$ state}, but
{\it definitely not a pure gluonium}.
%\end{itemize}
\section{SCALAR MIXING-OLOGY}
\subsection{Mixing below 1 GeV and nature of the $\sigma (1000)$ and $f_0(980)$}
We consider the two-component mixing scheme of the bare states $(\sigma_B,~S_2)$:
\beq
|f_0>\equiv -\sin\theta_s|\sigma_B>+\cos\theta_s|S_2>~,~~~~~~~~~~~~~
|\sigma>\equiv ~~~\cos\theta_s|\sigma_B>+\sin\theta_s|S_2>
\eeq
A sum rule analysis of the off-diagonal 2-point correlator \cite{MENES,SNB,SNG}:
\beq\label{offs}
\psi_{qg}^S(q^2) \equiv i \int d^4x ~e^{iqx}
\la 0\vert {\cal T}
\beta(\alpha_s)
G^2(x)\sum_{u,d,s}m_q\bar
qq~(0) \vert 0 \ra ,
\eeq
responsible for the mass-shift of the mixed
states gives a small {\it mass mixing angle} of about $15^0$, which has been confirmed by lattice calculations
using different unput for the masses \cite{WEIN2} and from the low-energy theorems based on Ward identities
of broken scale invariance \cite{ELLIS}, if one uses there the new input values \cite{SNB,SNI} of the quark and
gluon condensates. In order to have more complete discussions on the gluon content of the different states, one
should also determine the {\it decay mixing angle}.  In so doing, we use the predictions for
$\sigma_B, ~S_2\rar
\gamma\gamma$ obtained in the previous sections and the data
$\Gamma(f_0\rar\gamma\gamma)\approx 0.3$ keV. Then, we deduce  a {\it maximal decay mixing angle}
and the widths \cite{BRAMON2,SNB}:
\bea
|\theta_s| &\simeq& (40-45)^0~,\nnb\\
\Gamma (f_0\rar\pi^+\pi^-)&\leq& 134~ {\rm
MeV}~,~~~~~~~~~~~~~~~~~~~ g_{f_0K^+K^-}/g_{f_0\pi^+\pi^-}\approx 2~,\nnb\\
\Gamma (\sigma\rar\pi^+\pi^-)&\approx& (300-700)~ {\rm MeV}~,~~~~~~~~
\Gamma (\sigma\rar\gamma\gamma)\approx ~(0.2-0.5) ~{\rm keV}~.
\eea
The huge coupling of the $f_0$ to $\bar KK$ comes from the large mixing with the 
$\sigma$. For this reason, the $f_0$ can have a large singlet component, as also suggested from
independent analysis \cite{MONT,MINK}. Extending the previous $J/\psi \rar \gamma +X$ analysis into the
case of the $\phi$, one obtains the {\it new result} within this scheme \cite{SNU}:
\beq
Br [\phi\rar \gamma
+f_0(980)]\approx 1.3\times 10^{-4}~,
\eeq
in good agreement with the Novosibirsk data of $(1.93\pm 0.46\pm 0.5)\times 10^{-4}$.
\subsection{Mixing above 1 GeV  and nature of the $f_0(1.37)$,
$f_0(1.5)$, $f_0(1.6)$  and $f_0(1.7)$}
As already mentioned previously, this region is quite complicated due to the
proliferation of states.
Many scenarios have been proposed in the literature for trying to interpret this region
\cite{AMSLER,SNG,MONT,MINK,WEIN2}. However, one needs to clarify and to confirm the data \cite{GAST} for selecting
these different interpretations. We shall give below {\it some selection rules} which can already eliminate some of
these different schemes:\\
$\bl$ {\bf The $f_0(1.37)$} decay into $\sigma\sigma\rar (4\pi^0)_S$ signals mixings with the
$\sigma$, $\sigma'$ and $G$. \\
$\bl$ 
{\bf The $f_0(1.5)$ of Crystal Barrel and the $f_0(1.6)$ of
GAMS} are different objects:\\ 
%\begin{itemize}
$-$ Both states like to go into $\sigma\sigma$ and $\eta'\eta$, which signals a 
gluon component.\\
$ -$ The $f_0(1.5)$ coupled to $\pi\pi$ and $\bar KK$ signals a $\bar qq$ component which
may come from the $S'_2,~S_3$. Then, it can result from the $\bar qq$ mixings with the $\sigma,~\sigma'$ and
$G$,  like the
$f_0(1.37)$.\\  
$-$ The $f_0(1.6)$ couples weakly to $\pi\pi$ and $\bar KK$, while the ratio of its
$\eta\eta$ and $\eta'\eta$ is proportionnal to $1/\sin^2 \theta_P$. These features indicate that it is
an almost pure gluonium state, which can be identified with the $G$ in Table 1 obtained in the quenched
approximation (this approximation is expected to better at higher energies using $1/N_c$ arguments 
\cite{VEN,GAB}). \\
$\bl$ {\bf The $f_0(1.7)$} is the radial excitation $S'_3$ of the $S_3(\bar
ss)$ state (see section 7).
%\end{itemize}
\section{LIGHT $1^{-+}$ HYBRIDS}
The experimental situation has been discussed in \cite{CHUNG}. The sum rule analysis of the
spectrum is based on the 2-point correlator $\psi(q^2)_H$ associated to the hybrid currents:
\beq
{\cal O}^V_\mu (x)\equiv :g\bar \psi_i\lambda_a \gamma_\nu \psi_j G^{\mu\nu}_a:
~,~~~~~~~~~~~~~~~~~~~~~~~{\cal O}^A_\mu (x)\equiv :g\bar \psi_i\lambda_a 
\gamma_\nu \gamma_5\psi_j G^{\mu\nu}_a:
\eeq
which shows that the lowest state is the $1^{-+}$.
%\subsection{The $\tilde\rho(1^{-+})$}.
\\
$\bl$ An analysis of the $\tilde\rho(1^{-+})$ mass and coupling have been done in the past by different
groups \cite{HYG,HYG2}, where (unfortunately) the non-trivial QCD expressions were wrong leading to some
controversial predictions \cite{SNB}. The final correct QCD expression is given in
\cite{LNP}. The analysis has been extended recently taking into account the effect of a non-trivial $1/q^2$ term
in the OPE. One obtains the preliminary results \cite{CNP}:
\beq
M_{\tilde\rho}\approx 1.6~\mbox{GeV}~,~~~~~~~~f_{\tilde\rho}\approx (25\sim 50)~\mbox{MeV}~,~~~~~~~~
(M_{{\tilde\rho}'}\approx\sqrt{t_c})-M_{{\tilde\rho}}\approx 200
~\rm{MeV}~, 
\eeq
where the $\tilde\rho'$ is the radial excitation. One can consider this result as an improvment of the
available sum rule results ranging from 1.4 to 2.1 GeV, \cite{SNB}, though, we cannot absolutely exclude
the presence of the 1.4 GeV candidate \cite{CHUNG}.  
The
$\tilde\rho'$-$\tilde\rho$ mass-splitting is much smaller than
$M_{\rho'}-M_\rho\simeq 700$ MeV, and can signal a rich population of $1^{-+}$ states above 1.6 GeV. The hadronic
widths have been computed in
\cite{LNP,VIRON}, with the values:
\bea
\Gamma(\tilde\rho\rar \rho\pi)&\approx& 274~\mbox{MeV}~,~~~~~~~~~~~~~~~~
\Gamma(\tilde\rho\rar \gamma\pi)\approx 3~\mbox{MeV}~,\nnb\\
\Gamma(\tilde\rho\rar \eta'\pi)&\approx& 3~\mbox{MeV}~,~~~~~~~~~~~~~~~~~~~
\Gamma(\tilde\rho\rar \pi\pi,~\bar K K,~\eta_8\eta_8 )\approx {\cal O}(m^2_q)~.
\eea
$\bl$ These new \cite{CNP} and old \cite{SNB} values of
$(M_{\tilde\rho},~f_{\tilde\rho})$, indicate that the constraint \cite{HYG}:
 \beq\psi_H(0)\approx \frac{16\pi}{9}\alpha_s
\la \bar\psi\psi\ra^2~,
\eeq
is quite inaccurate as it underestimates the value of $\psi_H(0)\simeq 2f^2_{\tilde\rho}M^4_{\tilde\rho}$
by a factor 10. An independent measurement of this quantity, e.g. on the lattice,
can be useful.\\
$\bl$ One can measure the $SU(3)$ breakings and the mass of the $\tilde\phi(\bar ss)$
from the difference of the ratio of moments, which gives:
\beq
 M^2_{\tilde\phi}-M^2_{\tilde\rho}\simeq \frac{20}{3}\overline
m^2_s-\frac{160\pi^2}{9}m_s\la
\bar ss\ra \tau \approx 0.3 ~\mbox{GeV}^2~~~~\lrar~~~~ M_{\tilde\phi} \approx
1.7~\mbox{GeV}~,
\eeq
a value which is in the range of the lattice
results of $(1.7\sim 2.2)$ GeV \cite{LATH}.
\section{HEAVY $1^{-+}$ HYBRIDS }
%\subsection{Masses of the $\bar QQg$ hybrids}
The sum rule predictions for the masses are \cite{HYBH,HYBHQET}:
\beq
M_{\tilde{\psi}(\bar ccg)}\simeq 4.1~\rm{GeV}~,~~~~~~~~M_{\tilde{\Upsilon}(\bar bbg)}\simeq
10.6~\rm GeV~,~~~~~~~~M_{\tilde{B^*}(\bar bug)} \simeq 6.3 ~\rm GeV~,
\eeq
 where the two former agree (within the
10\% systematics of the two methods) with the available lattice values \cite{HYBLAT} of about 4.4 GeV [resp.
$11$ GeV], and are above the $\psi$-$\pi$ [resp. $\Upsilon$-$\pi$] thresholds. A check of these results is
in progress \cite{CNP}.
Using the
$1/M_b$-expansion sum rule, the
$\tilde{B^*}(\bar bug)$ decays are found to be \cite{HYBHQET}:
\beq
\Gamma (\tilde{B^*}\rar B_1\pi)\approx 250~\rm{ MeV}~,~~~~~~~~~~~~~~~~~
\Gamma (\tilde{B^*}\rar \sum _{X\neq B_1}X\pi)\approx 50~\mbox{ MeV}~.
\eeq
% A check of these results are under investigation in \cite{CNP}.
\section{CONCLUSIONS}
% AND ACKNOWLEDGEMENTS} 
There are some progresses in the long run study and search for the exotics. Before some
definite conclusions, one still needs improvments of the present data, and some lattice unquenched
estimates of the mixing angles and widths which should complement the QCD spectral sum rule
(QSSR) and low-energy theorem (LET) results. However, one can already notice that the
simple $\bar qq$-scheme combined with (large) mixings with gluonia (the $\sigma$,...) 
can explain the complex
$0^{++}$ data. It is a pleasure to thank the organizers of Hadron 99 for their
invitation to present this review in this exotic country !


\begin{thebibliography}{999}
\bibitem{GM}M. Gell-Mann, {\it Acta Phys. Aust. Suppl} {\bf 9} 
(1972) 733; H. Fritzsch and M. Gell-Mann, {\it XVI Int. Conf. 
High-Energy
Phys.}, Chicago, {\bf Vol 2} (1972) 135. 
%H. Fritzsch and P. Minkowski, {\it Nuovo Cim} {\bf A30} (1975) 393.
\bibitem{U1} E. Witten, {\it Nucl. Phys.}
{B156} (1979) 269; G. Veneziano, {\it Nucl. Phys.}
{B159} (1979) 213.
\bibitem{REVF} N. Isgur and J.E. Paton, {\it Phys. Rev.} {\bf D31} (1985)
1910; T. Barnes, F.E. Close, P.R. Page and E. Swanson,
{\it Phys. Rev.} {\bf D55} (1997) 4157 and references therein.
\bibitem{REVB}R.L. Jaffe, {\it Phys. Rev.} {\bf D15}(1977)267;
M. Chanowitz and S. Sharpe, {\it Nucl. Phys.}
{B222} (1983) 211; T. Barnes, F.E. Close and F. de Viron, {\it Nucl. Phys.}
{B224} (1983) 241.
\bibitem{REVP} S. Godfrey and N. Isgur, {\it Phys. Rev.} {\bf D32} (1985)
189; A. Le Yaounac et al., {\it Phys. Rev.} {\bf
D8} (1985) 2223 (1973); N. Tornquist, {\it Ann. Phys.} {\bf 123} (1979) 1.
\bibitem{REVG} E. Swanson, QCD 99
(Montpellier) and references therein. 
\bibitem{SVZ} M.A. Shifman, A.I. Vainshtein and V.I. Zakharov,
{\it Nucl. Phys.} {\bf B147} (1979) 385, 448.
\bibitem{SNB}S. Narison, Lecture Notes in
Physics, {\bf Vol. 26} (1989) ed. World Scientific.
\bibitem{ZAK} V.A. Zakharov, QCD 99
(Montpellier) and references therein. 
\bibitem{SNI} S. Narison, {\it Phys. Lett.} {\bf B300} (1993) 293; {\bf B361}
(1995) 121.
\bibitem{CNZ} K. Chetyrkin, S. Narison and V.A. Zakharov, {\it Nucl. Phys.}
{B550} (1999) 353.
\bibitem{TEPER} See e.g., M. Teper, hep-ph/9711299 (1997); hep-lat 9711011 (1997).
\bibitem{SNH} S. Narison, {\it Phys. Lett.} {\bf B387} (1996) 162.
\bibitem{NSVZ}V.A. Novikov et al, {\it Nucl. Phys.} {\bf B191} (1981) 301.
\bibitem{RAFAEL} R.A. Bertlmann, G. Launer and E. de Rafael, {\it Nucl. Phys.}
{\bf B250} (1985) 61; R.A. Bertlmann et al., {\it Z. Phys.} {\bf C39} (1988)
231 and references therein.
\bibitem{BNP} E. Braaten, S. Narison and A. Pich, {\it Nucl. Phys.} {\bf B 373} (1992) 581.
\bibitem{SNG} S. Narison, {\it Nucl. Phys.} {\bf B509} (1998) 312; {\it Nucl. Phys. (Proc.
Suppl.)} {\bf B64} (1998) 210.
\bibitem{LATT} G. Bali et al., {\it Phys. Lett.} {\bf B 309} (1994) 29; 
J. Sexton et al., 
{\it Nucl. Phys. (Proc. Suppl.)} {\bf B42} (1995) 279; C.J. Morningstar and M. Peardon,
hep-lat/9901004 (1999).
%and {\bf B23} (1991) 280.
\bibitem{WEST} G. West,  {\it  
QCD96},
Montpellier, {\it Nucl. Phys. (Proc. Suppl.)} {\bf B, A54} (1997).
\bibitem{SHORE}S. Narison, G. Shore and G. Veneziano, {\it Nucl. Phys.} {\bf B433} (1995)209;
{\bf B546} (1999)235.
\bibitem{PAK}S. Narison, N. Pak and N. Paver, {\it Phys. Lett.} {\bf B147}
(1984) 162.
\bibitem{LIPKIN}H. Lipkin, these proceedings and private communication.
\bibitem{BES} BES collaboration: C.Z. Yuan, QCD 99 (Montpellier); Z.G. Zhao, these proceedings.
\bibitem{BRAMON} E. Bagan, A. Bramon and S. Narison,
{\it Phys. Lett.} {\bf B196} (1987) 203.
\bibitem{SHIF}T.M. Aliev and M.A. Shifman, {\it Phys. Lett.} {\bf B112}
(1982) 401.
\bibitem{BUGG}D. Bugg, these proceedings and private comunication.
\bibitem{VEN}S. Narison and G. Veneziano, {\it Int. J. Mod. Phys} 
{\bf A4, 11} (1989) 2751.
\bibitem{MONT} For a review, see e.g. L. Montanet, QCD 99 (Montpellier) and private
communications.
\bibitem{MINK}P. Minkowski and W. Ochs, {\it Eur. Phys. J} {\bf C9} (1999) 283
and private communications.
\bibitem{ANIS}V. Anisovich et al., {\it Phys. Lett.} {\bf B413}
(1997) 476.
\bibitem{CHAN}M.S. Chanowitz, Workshop on $\gamma\gamma$-collisions (Israel 1988).
\bibitem{MALT}K. Maltman, these proceedings.
\bibitem{BRAMON2}A. Bramon and S. Narison, {\it Mod. Phys. Lett.} 
{\bf A4} (1989) 1113.
\bibitem{SN4}S. Narison, {\it Phys. Lett.} {\bf B175} (1986) 88.
\bibitem{LEE}W. Lee and D. Weingarten, hep-lat/9608071.
\bibitem{SING} A. Singovski for the WA102 collaboration, QCD 99 (Montpellier); these proceedings.
\bibitem{MENES}G. Mennessier, S. Narison and N. Paver, {\it Phys. Lett.} {\bf B158}
(1985) 153.
\bibitem{WEIN2}W. Lee and D. Weingarten, hep-lat/9805029.
\bibitem{ELLIS}J. Ellis, H. Fujii and D. Kharzeev, hep-ph/9909322.
\bibitem{SNU}S. Narison (unpublished).
\bibitem{AMSLER}C.Amsler and F.E. Close, {\it Phys. Rev.} {\bf D53}
(1996) 296.
\bibitem{GAST}U. Gastaldi, QCD 99 (Montpellier) and private communications.
\bibitem{GAB}G. Veneziano, private communications.
\bibitem{CHUNG}S.U. Chung, QCD 99 (Montpellier); these proceedings; P. Eugenio, these proceedings.
\bibitem{HYG}I.I Balitsky, D.I. D'Yakonov and A.V. Yung, {\it Phys. Lett.} {\bf B112}
(1982) 71.
\bibitem{HYG2} J. Govaerts et al., {\it Nucl. Phys.} {\bf B248} (1984) 1; J.I. Latorre et al.,
{\it Phys. Lett.} {\bf B147} (1984) 169.
\bibitem{LNP}J.I. Latorre, S. Narison and P. Pascual, {\it Z. Phys.} {\bf C34} (1987)
347.
\bibitem{CNP}K. Chetyrkin, S. Narison and A. Pivovarov (to appear).
\bibitem{VIRON} J. Govaerts and F. de Viron, {Phys. Rev. Lett.} {\bf 53} (1984) 2207.
\bibitem{LATH}C. Bernard et al., hep-lat/9809087; P. Lacock et al. hep-lat/9809022.
\bibitem{HYBH}J. Govaerts et al., {\it Nucl. Phys.} {\bf B258} (1985) 215; {\bf B262} (1985) 575;
{\bf B284} (1987) 674.
\bibitem{HYBHQET} S-L. Zhu, hep-ph/9812469; T. Huang, H. Jin and A. Zhang, hep-ph/9812430.
\bibitem{HYBLAT}T. Manke, QCD 99 (Montpellier).
\end{thebibliography}
\end{document}